\documentclass[twocolumn,letterpaper,aps,prc,longbibliography,superscriptaddress,nofootinbib,floatfix]{revtex4-1}

\usepackage{caption}
\usepackage{subcaption}
\usepackage{graphicx}	% Include figure files
\usepackage{xspace}     % Include xspace
\usepackage{amsmath}

\begin{document}

\title{Verification of spent nuclear fuel in sealed dry storage casks via measurements of cosmic ray muon scattering}

\newcommand{\losalamos}{Los Alamos National Laboratory, Los Alamos, New Mexico 87545, USA}
\newcommand{\newmex}{University of New Mexico, Albuquerque, New Mexico 87131, USA}
\newcommand{\inl}{Idaho National Laboratory, Idaho Falls, Idaho 83415, USA}

\affiliation{\losalamos}
\affiliation{\newmex}
\affiliation{\inl}

\author{J. M. Durham} \email[Corresponding author: ]{durham@lanl.gov} \affiliation{\losalamos} 
\author{D. Poulson} \affiliation{\losalamos} \affiliation{\newmex}
\author{J. Bacon} \affiliation{\losalamos}
\author{D. L. Chichester} \affiliation{\inl}
\author{E. Guardincerri} \affiliation{\losalamos}
\author{C. L. Morris} \affiliation{\losalamos}
\author{K. Plaud-Ramos} \affiliation{\losalamos}
\author{W. Schwendiman} \affiliation{\inl}
\author{J. D. Tolman} \affiliation{\inl}
\author{P. Winston} \affiliation{\inl}
\date{\today}

\begin{abstract}

Most of the plutonium in the world resides inside spent nuclear reactor fuel rods.  This high-level radioactive waste is commonly held in long-term storage within large, heavily shielded casks.  Currently, international nuclear safeguards inspectors have no stand-alone method of verifying the amount of reactor fuel stored within a sealed cask. Here we demonstrate experimentally that measurements of the scattering angles of cosmic ray muons which pass through a storage cask can be used to determine if spent fuel assemblies are missing without opening the cask.  This application of technology and methods commonly used in high-energy particle physics provides a potential solution to this long-standing problem in international nuclear safeguards.  

\end{abstract}

\maketitle

\section{Introduction}

Plutonium is produced in all uranium-fueled nuclear reactors following neutron capture on $^{238}$U nuclei and two beta decays, primarily through the reactions

\begin{equation}
^{238}\textbf{U} + n {\xrightarrow{~~~~}} ^{239}\textbf{U} {\xrightarrow{~\beta^{-}~}} {^{239}\textbf{Np}} {\xrightarrow{~\beta^{-}~}}  {^{239}\textbf{Pu}}. 
\end{equation}

As reactor fuel ages, the concentration of fission products in the fuel grows, reducing the reactor's neutron flux and overall efficiency. Typically, when spent fuel is removed from power reactors, $\sim1\%$ of the initial $^{238}$U nuclei have been converted to various plutonium isotopes, with relative concentrations that depend on the total neutron flux the fuel has been exposed to and the age of the fuel.  After removal from the reactor core, the plutonium can be recovered and purified by chemical reprocessing of the spent fuel.  It can then be used in radioisotope thermoelectric generators \cite{RTG,RTG2}, recycled for further use as reactor fuel \cite{MOX}, or used as fissile material in nuclear explosives.

Most nations in the world are signatories to the Treaty on the Non-Proliferation of Nuclear Weapons (NPT) \cite{NPT}. The NPT has three key pillars, which specify treaty member's requirements regarding nonproliferation, the peaceful uses of nuclear energy, and disarmament. In relation to nonproliferation, the nuclear-weapon states\footnote{defined in the NPT as nations which manufactured and detonated a nuclear explosive device prior to 1 January 1967: China, France, Russia, the United Kingdom, and the United States} pledge not to directly or indirectly facilitate the transfer of nuclear weapons to non-weapon states, or to facilitate their acquisition or control of such weapons. The non-weapon states pledge not to receive, manufacture, or otherwise acquire nuclear weapons. In this context, the non-weapon states agree to accept safeguards by the International Atomic Energy Agency (IAEA), which shall be applied on all fissionable material within the state or under its control for the purpose of verifying compliance with the treaty. The IAEA considers member state declarations, analyzes open-source data and information from third parties, and conducts inspections of nuclear facilities to draw conclusions about activities in states under nuclear safeguards.

Inspectors have an array of technologies at their disposal to examine materials at various stages of the nuclear fuel cycle \cite{IAEA_tech, Safeguards}.  Many sites which are under safeguards have cameras, allowing inspectors to remotely monitor reactor fuel loading and unloading, and portal monitors on facility exits which can detect spent fuel assemblies being moved out of the reactor building \cite{IAEA_bulletin, IAEA_bulletin2}.  Following removal from the reactor core, spent fuel usually goes directly into a cooling pool, where it sits for several years as short-lived fission products decay. Here inspectors can verify the presence and number of fuel rods by observing Cherenkov radiation produced in the water by electrons which have Compton scattered from decay photons produced by fission products \cite{DCVD}, or with neutron and gamma ray measurements \cite{SMOPY}.  

After removal from the pool, spent fuel is typically placed in long-term storage within large, heavily shielded containers called dry storage casks \cite{NUREG}.  A typical cask for fuel from pressurized light-water reactors is a right circular cylinder of diameter $\sim$3 m and height $\sim$5 m, and consists of a central basket which holds 20-30 spent fuel assemblies, surrounded by cylindrical layers of neutron and gamma-ray shielding.  IAEA inspectors maintain accountability of the cask contents by applying tamper-indicating seals to the lids of the casks, which are checked during periodic inspections.   

Once filled, casks are typically stored outdoors at reactors or dedicated storage facilities.  At these sites, the cask seals are exposed to the elements, where they may corrode, or can be damaged during cask handling.  If a seal fails, accountability of the fuel in the cask is lost; to reassert the IAEA's knowledge of its contents the cask must be returned to a spent fuel pool where it can be safely opened and re-verified.  The long procedure for transporting and opening the cask is costly, as well as disruptive to operations at the nuclear facility under inspection.  Therefore, an $in$ $situ$ means for determining if the cask's contents are intact is a necessary tool for safeguards inspectors \cite{LTRD, STR382}.

Conventional active radiography with neutrons or photons is not feasible due to the heavy shielding that is used to contain the radiation emitted by the fuel, as well as self-shielding from the fuel assemblies themselves, which can total over 100 radiation lengths for typical casks \cite{CHATZI1}.  Previous work showed that measurements of radiation which escapes the cask are capable of proving that the cask has radioactive contents, but the scattered radiation which emerges does not carry sufficient radiographic information to determine if individual assemblies are missing \cite{Ziock}.  Alternative methods of measuring a cask's radiation ``fingerprint" for subsequent re-verification at later dates \cite{Ziock2, DSVD} must correct for decays of fission products in the fuel, and for large variations in background between the initial measurement at the reactor site and later measurements at a spent fuel storage installation.  Since this requires pre-existing measurements of a cask's condition, as well as information on fuel burnup from the facility operator, it is not an independent, stand-alone method for re-verification of existing casks.  Antineutrino monitoring has also been studied as a method for detection of missing fuel \cite{neutrinos}, however this requires detectors with active masses on the order of $\sim$10 tons, which presents a deployment challenge,  and requires counting time on the order of $\sim$1 year.   There are many sites in Europe and Asia where spent fuel is monitored under safeguards, but this current inability to determine the content of dry storage casks also presents a potential challenge to full implementation of the Joint Comprehensive Plan of Action (JCPOA) \cite{JCPOA}, an international agreement which includes requirements that Iran remove all spent reactor fuel from the country.  The highly radioactive spent fuel from the Arak heavy-water reactor must be stored and shipped in heavily shielded containers, which may potentially require verification prior to shipment.

There has been much work in simulation and on laboratory test objects to show that highly penetrating cosmic-ray muons are potentially useful for interrogating encapsulated nuclear waste \cite{Gustafsson, Jonkmans, Furlan, Liao, Clarkson1, Ambrosino, Clarkson2, CHATZI1, Frazao, Chatzi2, Poulson, Boniface, Checchia, Liu}. A previous measurement by our group showed that muon scattering is sensitive to the fuel content of a cask, but that data set only covered a small portion of a cask and had limited discriminatory power \cite{Durham1}.  Here, we show experimentally that muon scattering radiography is sensitive to the removal of multiple fuel bundles from a dry storage cask with high confidence ($>5\sigma$), and demonstrate a potential sensitivity to the removal of a single bundle at the 2.3$\sigma$ level in a statistics-limited measurement.  This potentially represents a new method for inspectors to verify the content of a dry storage cask, and can thereby solve a long-standing problem in nuclear safeguards.

\section{Method}

Highly energetic nuclei produced in astrophysical processes are continually interacting in the upper atmosphere, producing showers containing secondary charged mesons that subsequently decay to muons.  These cosmic ray muons arrive on the surface of the Earth at a rate of $\sim10^4/$m$^2/$min with a broad energy distribution that has a mean of $\sim$4 GeV (see \cite{PDG} for a review). The muon flux is roughly proportional to $cos^{2}\theta_{z},$ where $\theta_z$ is the angle from the zenith. Muons are the most abundant charged particle found at the Earth's surface.

The unique properties of muons enable them to penetrate large, dense objects and provide radiographic information on the object's internal structure, through measurements of muon attenuation \cite{George, Alvarez,volcano,DIAPHANE}, scattering \cite{Nature, Pried, Schultz}, and associated secondary particle production \cite{MIF}.  Since muons are charged leptons, they do not lose energy through hadronic interactions, but only experience electromagnetic energy loss and multiple Coulomb scattering as they pass through material.  The muon's relatively large mass of 105 MeV/$c^{2}$ suppresses radiative energy loss from emission of bremsstrahlung photons, and cosmic ray muons relevant for imaging fall in the momentum range where energy loss per unit areal density is near the Bethe-Bloch minimum of $-<dE/dx> \approx$ 1 MeV/g/cm$^2$.  The scattering angles relativistic muons undergo when passing through matter are dependent on the radiation length $X_{0}$ of the material encountered, and can be described by a Gaussian with a width approximated by  

\begin{equation}
\sigma = \frac{13.6 \mbox{MeV}}{\beta c p} \sqrt{L/X_{0}},
\end{equation}

where $\beta c$ and $p$ are the muon's velocity and momentum, and $L/X_{0}$ is the number of radiation lengths the muon passes through \cite{PDG}.  Since $X_{0}$ decreases rapidly with increasing atomic number $Z$, muon scattering is especially sensitive to the presence of dense, high-$Z$ materials.  These properties allow muons to penetrate low-$Z$ shielding (a cask's steel or concrete body), interact with high-$Z$ material (uranium fuel) and exit the shielding carrying radiographic information about the internal structure.

\begin{figure}[h]
	\centering
		 \includegraphics[width=0.5\textwidth]{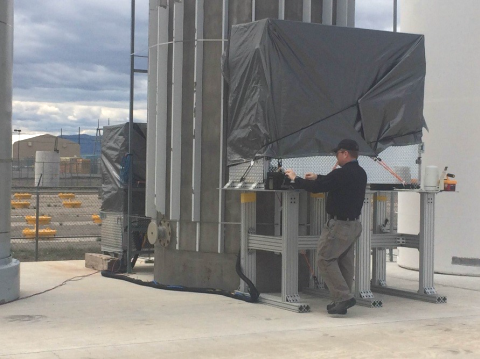}	
	\caption{The two muon trackers around the MC-10 cask. One tracker is elevated by 1.2 m relative to the other to sample the muon flux at smaller zenith angles.  Both are housed inside weatherproof containers.  One of the lifting trunnions that is used as a reference point to align the detectors with the fuel basket is visible on the bottom left of the cask body. }
	\label{fig:cask_photo}
\end{figure}

\begin{figure}[h]
	\centering
		 \includegraphics[width=0.45\textwidth]{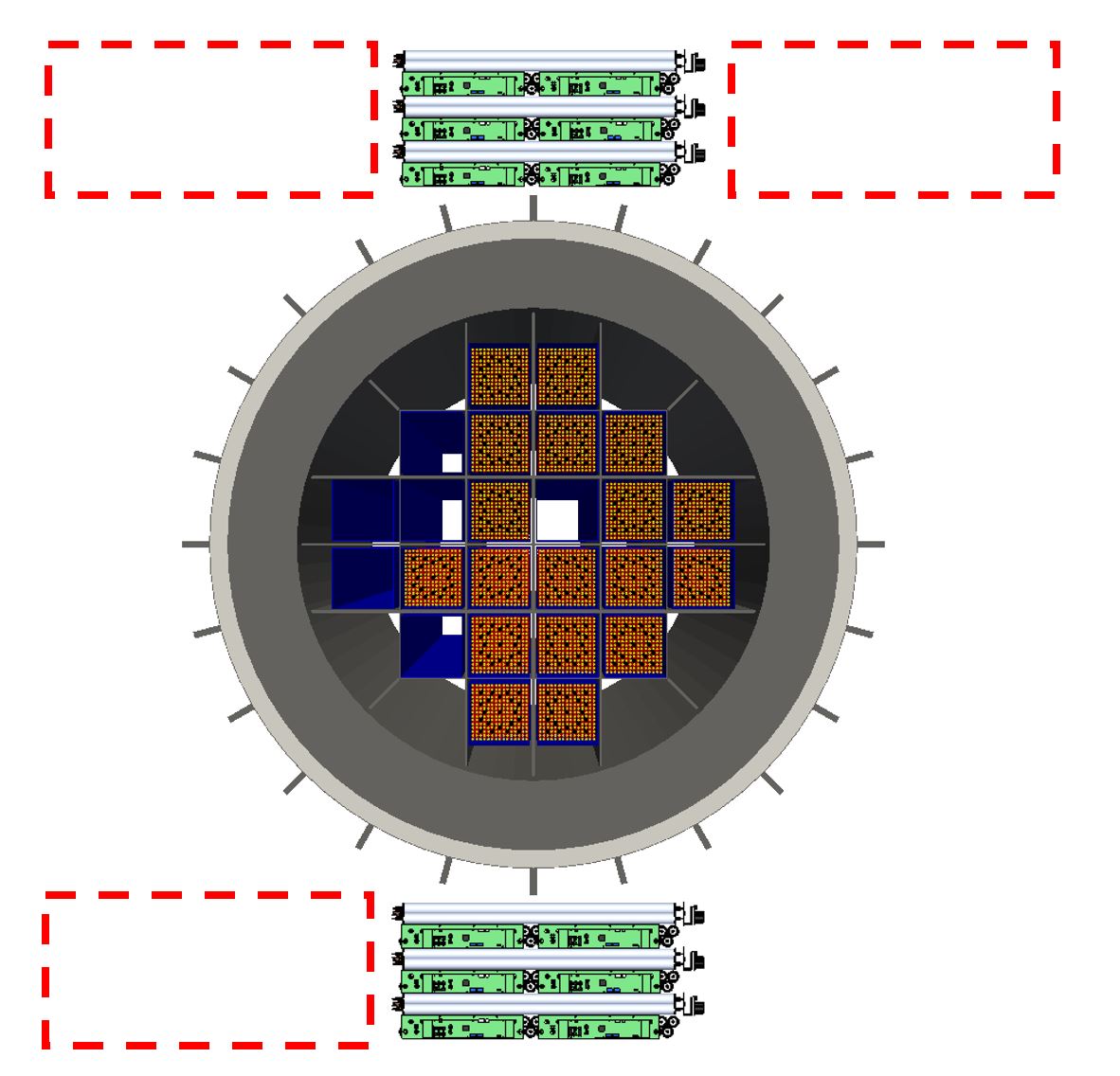}
	\caption{A diagram showing the cask loading configuration and the detector positions during the measurements as viewed from above. Muons moving between the two detectors pass through columns in the fuel basket containing (from left to right) zero, one, six, five, four, and two fuel assemblies. Data was recorded with the detector on the lower right, but winds at the testing site shook the detector during this portion of the measurement, rendering the data unusable for analysis.} 
	\label{fig:cask_load}
\end{figure}

In 2016, two muon tracking detectors were placed around a Westinghouse MC-10 cask at Idaho National Laboratory (see Fig. \ref{fig:cask_photo}) \cite{MC10}.  The MC-10 cask has an external diameter of 2.7 m and a height of 4.8 m.  Inside the steel skin of the cask is a cylindrical layer of BISCO NS-3 neutron shielding and a 25 cm thick layer of steel gamma shielding, which surrounds the fuel basket.  The basket itself has slots for 24 PWR fuel assemblies, and is made of aluminum with borated plates to control criticality.  The fuel assemblies inside the cask are standard Westinghouse 15$\times$15 rod bundles that were removed from a power reactor in the 1980s. Each assembly is 21$\times$21 cm$^2$ in cross section and 4 m in height, with a burnup in the range of 3$\times10^4$ MWday. This particular cask was chosen as a test object because it is only partially loaded, with 18 out of 24 possible fuel positions filled.  A diagram of the cask loading is shown in Fig. \ref{fig:cask_load}.  Lifting trunnions on the outside of the cask serve as a fiducial reference point for alignment of the muon detectors with the fuel basket.

\begin{figure}[h]
	\centering
		 \includegraphics[width=0.5\textwidth]{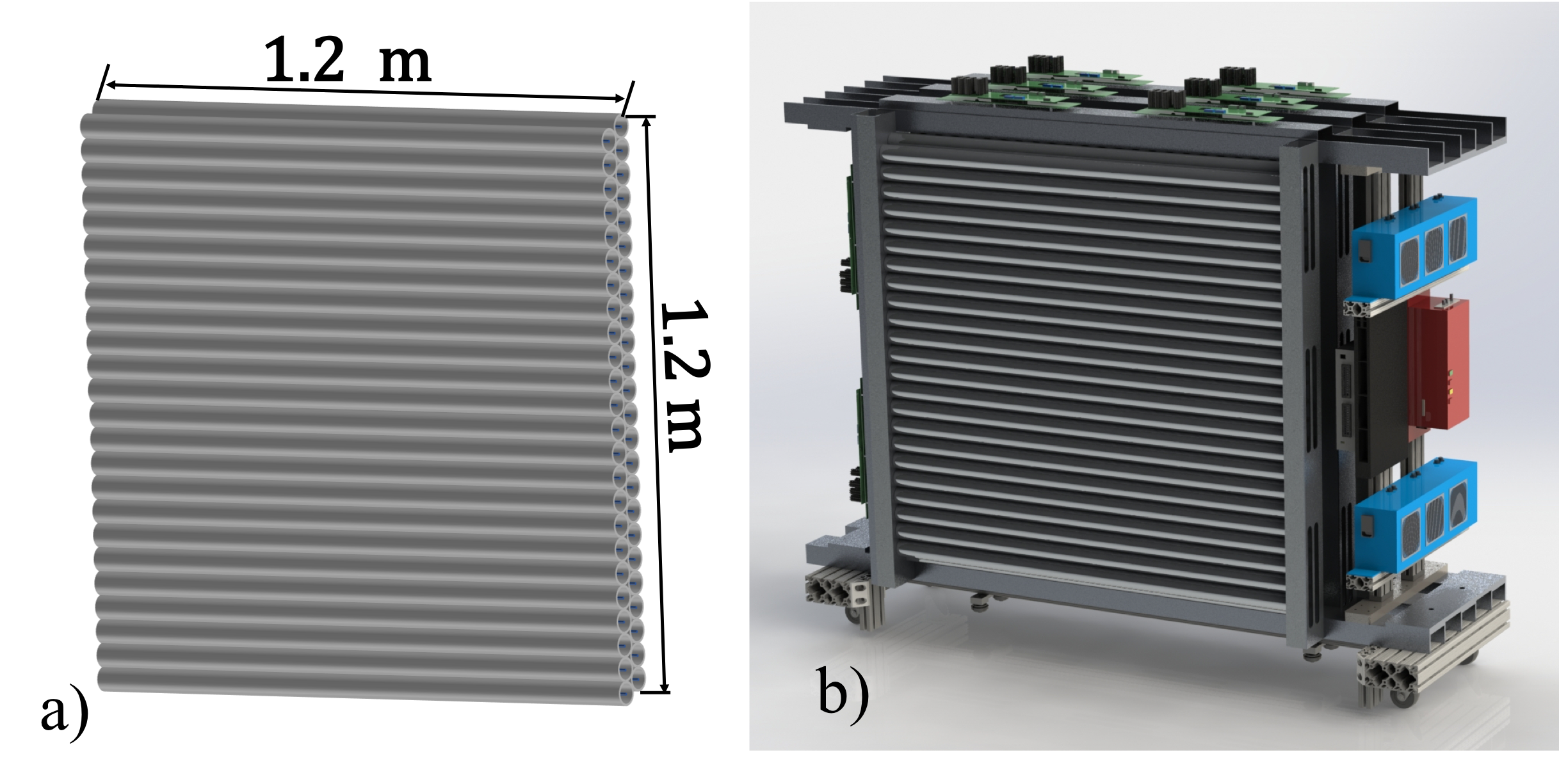}
	\caption{ a) One of the horizontal double-layers of drift tubes, which gives a position where the muon crossed the detector along the vertical direction, but provides no information on the horizontal position where there is no segmentation.  In the complete muon tracker (b), an identical double-layer that is oriented with tube segmentation along the horizontal direction is placed behind this one.  This pattern repeats three times to give a total of six double-layers (three horizontal and three vertical) in each detector.  The blue and red boxes contain low and high voltage supplies.}
	\label{fig:MMT}
\end{figure}

Two identical muon tracking detectors were placed on opposite sides of the cask to measure the trajectories of muons before and after passing through the cask, with one detector elevated relative to the other by 1.2 m in order to sample the higher muon flux at smaller $\theta_{z}$.  Each tracker consists of six double-layers of 24 aluminum drift tubes, which are a typical technology used for tracking cosmic-ray muons \cite{Toshiba} and muons produced at collider experiments \cite{ATLAS, LHCb, PANDA} .  The tubes each have an aluminum wall with thickness of 0.89 mm  and are 5 cm in diameter and 1.2 m in length, and are filled with a 47.5/42.5/7.5/2.5 mixture of Ar/CF$_4$/C$_2$H$_6$/He at 1 bar and permanently sealed.  Muons passing through the tubes ionize the gas, and the ionization is amplified through a gas avalanche process and collected on 30 $\mu$m gold-plated tungsten wires held at 2.6 kV which run down the center of each tube.  Fits to the patterns of hits in the tubes give muon trajectories.  A comparison of the ingoing and outgoing trajectories gives the scattering angle the muon underwent while passing through the cask.  A drawing of one double-layer and a complete muon tracking detector is shown in Fig. \ref{fig:MMT}.

%
%\begin{figure}[h]
%	\centering
%		 \includegraphics[trim={8cm 0 9cm 0},clip,width=0.45\textwidth]{MMT}
%	\caption{A drawing of one muon tracking detector.  One of the horizontal layers of 24 drift tubes is visible on the front of the detector.  The next layer is identical but offset vertically by one tube radius.  The following two layers are identical to the first two but oriented vertically.  This pattern repeats three times to give a total of 12 layers (6 horizontal and 6 vertical) in each detector.  The blue and red boxes contain low and high voltage supplies.}
%	\label{fig:MMT}
%\end{figure} 

Despite the heavy shielding, there is still a radiation dose rate of $\sim$10 mrem/hr from neutrons and an additional $\sim$10 mrem/hr from gamma rays at the cask's outer surface, where the muon trackers are positioned.  Compton scattered electrons or protons from neutron scattering can produce hits in single tubes, which will degrade muon tracking performance if they are included in the track fitting algorithm.  To reduce this background, a trigger was introduced which requires hits in neighboring tubes within a time window of 600 ns in order for those hits to be considered in the tracking.  Given the maximum drift time of $\sim 1 \mu$s and the fact that $6.8\%$ of the drift tubes are non functional, this coincidence requirement reduces overall tracking efficiency by $\sim$50$\%$, but does allow muons to be tracked in this radiation environment.

Since the size of the detectors is small relative to the size of the cask, both the upper and lower detectors were placed at 3 different horizontal positions (for a total of 9 measurement configurations) in order to record scattering data across the entire cask.  At each position, a rough alignment of the detectors was performed using measurements taken by hand around the cask.  A fine alignment was later performed in software using the muon tracks themselves, with techniques that are commonly used to align charged particle tracking detectors at accelerator-based experiments.  Since there is no preferred direction for scattering of cosmic ray muons, the scattering angle in the plane perpendicular to the muons direction is zero when averaged over an ensemble of muon tracks.  Using the measured incoming and outgoing muon track vectors, an optimal rotation matrix to align the two directional vector sets was constructed using a least squares method described in \cite{Sorkine}. This rotation matrix was then applied to all of the muon tracks in a single data set. An optimal detector translation was then found by constructing the residual value

\begin{equation}
\begin{split}
S^2 \equiv \sum_{j=1}^m [(\vec{V_1}_j + \vec{O} - \vec{P_{21}}_j)^T(\vec{V_1}_j + \vec{O} - \vec{P_{21}}_j)\\ +  (\vec{V_2}_j - \vec{P_{12}}_j)^T(\vec{V_2}_j - \vec{P_{12}}_j)]
\end{split}
\end{equation}

where $\vec{V_i}$ is the position of a muon track in the $i$ detector and $\vec{P_{ik}}$ is the projection of a track in detector $i$ to detector $k$, and $\vec{O}$ is the translational offset vector. The sum is performed over all muon tracks in a single measurement configuration.  The residual $S^2$ was then minimized with respect to the offset vector using least squares.  Typically, the rotational corrections applied are on the order of a few milliradians and 1-2 cm, with remaining residuals that are less than 0.1 mrad. 
 
With both detectors at the edge of the cask (on the far left side of Fig. \ref{fig:cask_load}), the viewing area is centered on the 25 cm steel gamma ray shield, and the fuel content of the cask does not affect the measured scattering.  This provides a calibration data for the measurement as only the known composition of the cask body itself affects the muon scattering measured here.  Data was collected in each position for $\sim$10 days, with samples sizes ranging from 4$\times$10$^4$ to 9$\times$10$^4$ muon tracks in each configuration.  Unfortunately, strong winds at the outdoor measurement site shook the lower detector during the period when one side of the cask was being measured, which introduced artifacts into the data that cannot be corrected for with our track-based alignment procedure.  Therefore, we do not consider the data taken in this position (covering the rightmost portion of the cask) in our analysis.

\section{Data Analysis and Discussion}

As muons pass through the cask, the amount of scattering they undergo is dependent on the path lengths of cask shielding material and fuel encountered along their trajectory.  Since the amount of fuel between the detectors varies moving horizontally across the cask, the scattering largely depends on the muon's horizontal starting position and azimuthal angle.  Given that the fuel assemblies and body of the cask are approximately homogeneous in the vertical direction, and the detectors sample a relatively small range in zenith angle ($\sim 20^\circ$) , the zenith angle of the incoming muon has relatively little effect on the path length sampled, and therefore scattering angle.  We expect muon scattering to be sensitive to changes in fuel content as we move horizontally across the cask.

\begin{figure}[h]
	\centering
		 \includegraphics[width=0.45\textwidth]{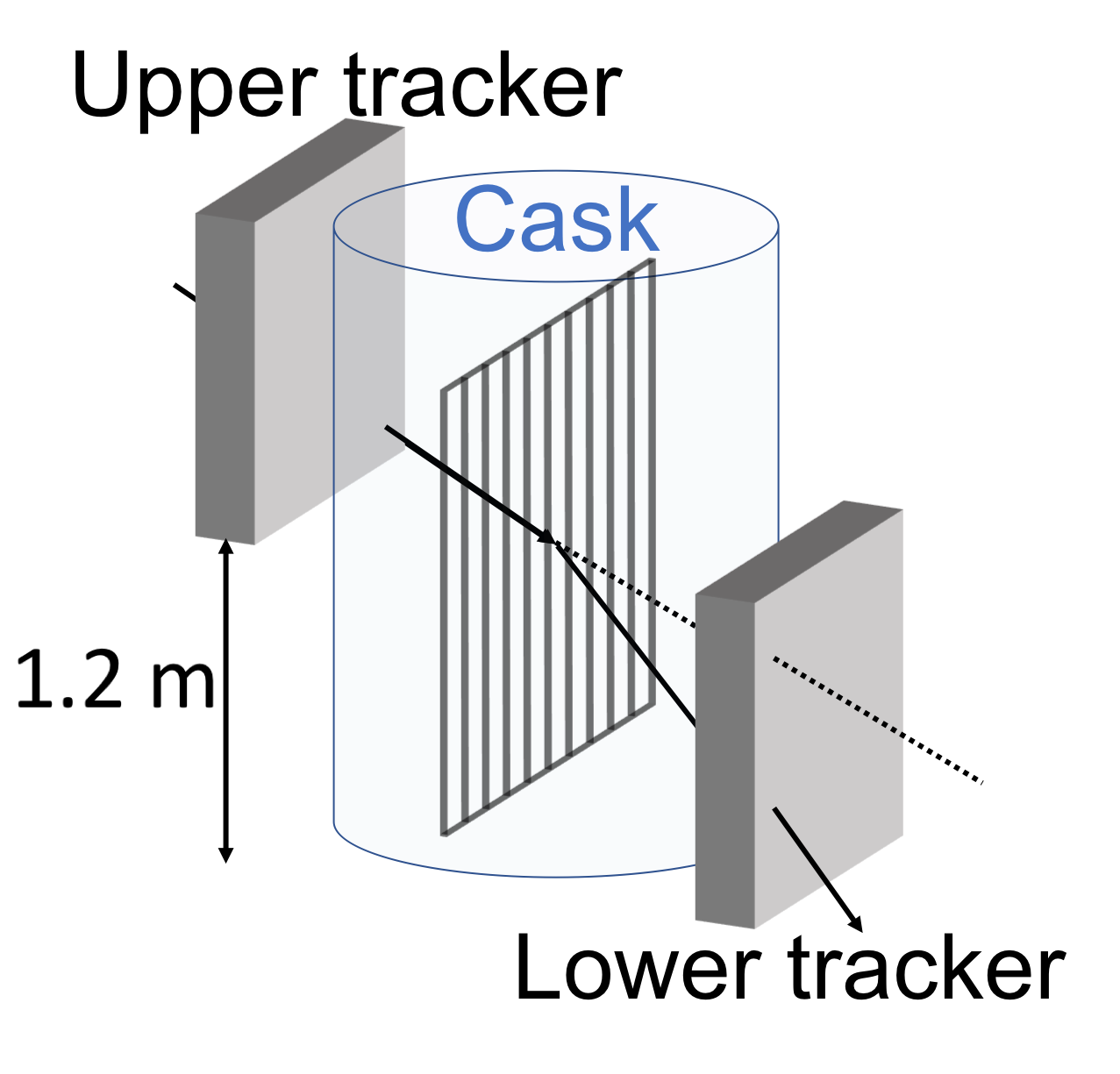}
	\caption{An illustration of the analysis procedure (not to scale).  Tracks measured in the elevated upper detector are projected to an array of voxels at the center of the cask.  Each voxel has a corresponding histogram where scattering angles of muons which pass through it are collected.}
	\label{fig:ana}
\end{figure}

Recognizing this, tracks in the upper detector are projected to the plane centered between the two detectors, which is divided into voxels with dimensions of 2 cm in the horizontal direction, 4 cm in the direction between the two detector, and 1.2 m (the size of the detectors, i.e. no division) in the vertical direction along the length of the fuel rods.  Each voxel has a corresponding histogram.  The histogram corresponding to each voxel the track passes through is filled with the absolute value of the muon scattering angle, weighted by the path length the muon traversed through that voxel.  Since the precise muon track within the cask cannot be measured, there is some uncertainty as to the exact position of the muon at the center of the cask, so the neighboring voxels are also filled with the same value.  See Fig. \ref{fig:ana} for an illustration of this procedure.

\begin{figure}[h]
	\centering
		 \includegraphics[width=0.45\textwidth]{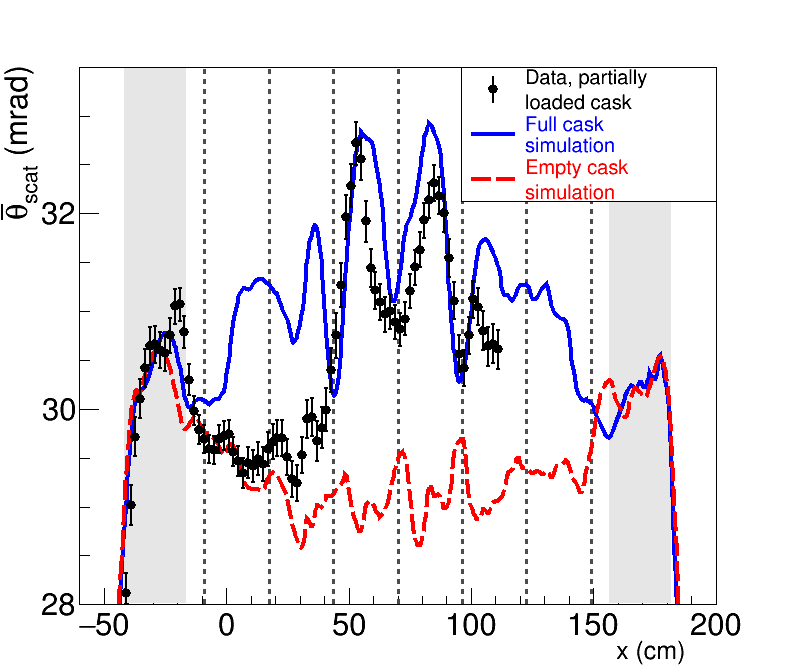}
	\caption{The average muon scattering angle as a function of horizontal position across the cask.  The gray shaded areas indicate the inner and outer boundaries of the 25 cm steel shielding around the fuel, and the boundaries of the columns in the fuel basket are denoted by dashed lines. A GEANT4 simulation of muon scattering in a full (empty) cask is shown by the solid blue (dashed red) line. }
	\label{fig:fig}
\end{figure}

The average value of the scattering angle in each voxel $\bar{\theta}_{scat}$ as a function of the horizontal coordinate across the cask is shown in Fig. \ref{fig:fig}, with error bars showing the statistical uncertainty.  The shaded areas represent the location of the steel cylindrical gamma shield, and dashed lines denote the boundaries of columns in the fuel basket. This metric is related to the number of radiation lengths the muons traverse while passing from one detector, through the cask, into the other detector, and is therefore sensitive to the amount of fuel in each column of the fuel basket.

The solid blue line and dashed red line on Fig. \ref{fig:fig} show $\bar{\theta}_{scat}$ from a GEANT4 simulation \cite{GEANT4} of muon scattering in a full and empty cask, respectively.  The 10$^7$ simulated muons used in each configuration were generated with angular and momentum distributions described in \cite{Poulson}, and the simulated detectors were tuned to have an active area that matched the instrument used for the measurement.  The measured data recorded in each of the nine detector configurations contains a different amount of muon tracks due to differences in detector solid angle and the time spent in each configuration, which introduces a bias towards the positions with higher counts when combining the voxel histograms from each position.  To account for this in simulation, the voxel histograms from each individual simulated detector configuration were filled with a weight corresponding to the number of muons recorded in that configuration in data relative to the entire actual data set.  Additionally, the coincidence requirement between neighboring tubes that is necessary to reconstruct muon tracks in the radiation field around the cask amplifies the effect of dead tubes in the detector, and introduces non-uniform angular biases in the detector active area.  These are accounted for in simulation by masking dead areas in the simulated detectors corresponding to holes in the actual detector acceptance.  Thus, while the cask body is a symmetric object, the simulation reflects asymmetries in the measured scattering due to detector acceptance effects and aberration due to different statistical samples in the various detector configurations. The simulated curves are normalized by a factor of 0.89$\pm$0.01 to match the data over the range inside the cask steel shielding, where contributions from fuel are negligible.

In the leftmost shaded area, a peak is seen due to scattering in the 25 cm steel gamma shielding surrounding the basket.  The precision on features in the first two basket columns are limited by statistical uncertainties, but there is evidence for a slight dip in the first column which contains no fuel as opposed to the two assemblies that would be present in a full cask. The data here is consistent with the expectation for an empty cask from the simulation. The second column shows a small peak corresponding to the single fuel assembly, and displays an average scattering magnitude between the expectations for a full and empty cask, indicating that this column is partially filled with fuel instead of containing four assemblies.  The third column in the fuel basket, which is fully loaded with six bundles, shows a clear peak with an amplitude that is consistent with the simulation of a full cask, and is larger than the next two columns containing five and four bundles, respectively.  The fourth column, which contains only five fuel assemblies instead of the full six, shows a peak with an amplitude smaller than the expectation for a fully populated row, which is indicative of the single missing fuel bundle.  The next row, which is fully loaded with four bundles, displays some disagreement with both the full and empty simulations, although the limited data on this row precludes strong conclusions about its fuel content.   

As previously discussed, data was recorded on the rightmost portion of the fuel basket, but during this part of the measurement the detectors were insufficiently secured against wind at the outdoor testing site.  The wind shook the detectors and therefore introduced artifacts in the measured scattering angles due to detector movement.  Therefore, this data was discarded and no statement can be made about the fuel content of the rightmost column.  The problem with detector movement will be corrected in future measurements with more secure footing.

 \begin{figure}[h]
 	\centering
 		 \includegraphics[width=0.45\textwidth]{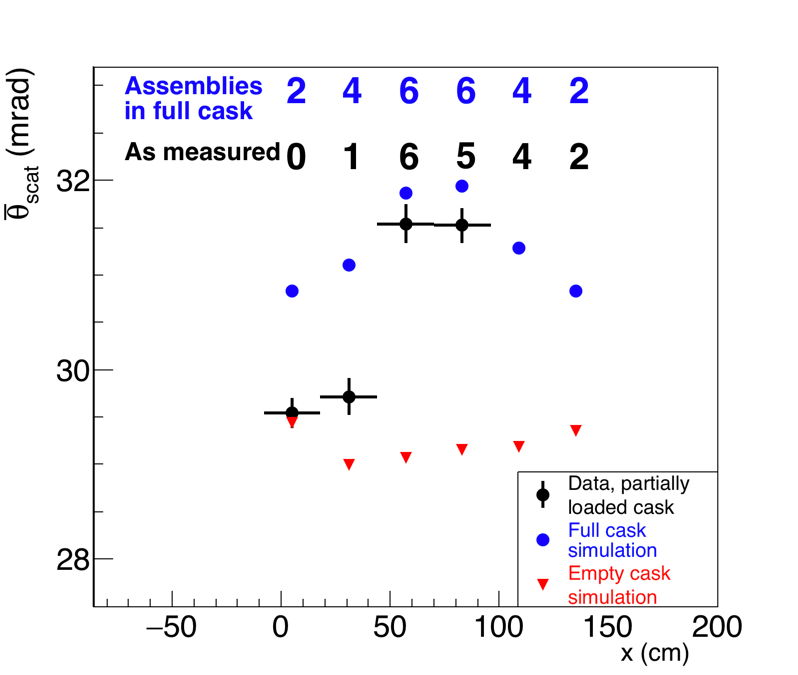}
 	\caption{The muon scattering signal, averaged across the positions of each column in the fuel basket, for data and simulations of a full and empty cask. The number of assemblies in each column for the full cask and the configuration as measured are indicated in text.}
 	\label{fig:fig4}
 \end{figure} 

For a quantitative comparison, the scattering signal from Fig. \ref{fig:fig} is averaged across the boundaries of each fuel column for which there is complete data and shown in Fig. \ref{fig:fig4}, along with the data from the simulations.  The vertical error bars indicate the statistical uncertainty and the horizontal bars show the range over which the averaging is done for each data point.  The simulation of the full cask shows a maximum scattering at the center of the cask, where muons pass through the most fuel.  The asymmetries between the sides of the cask with the same fuel content are due to the non-uniform angular acceptance of the detectors and the weights that are applied to each simulated measurement position to accurately reflect the instrument's active area and measurement times.  The empty cask simulation shows minimal scattering near the center of the cask, where the trajectories of muons moving from one detector to the other are nearly perpendicular to the steel gamma shielding and sample the least path length through this material.  On the edges of the cask, where the path length through the curving wall of the 25 cm gamma shielding in the cask is a maximum, the empty cask simulation shows increased scattering.

We see that the data from the first column, containing no fuel, is consistent with expectations for an empty cask.  The second column shows a scattering signal that is above the expectation for an empty cask, but is lower than the expected scattering in a full cask by 7$\sigma$ due to the three missing bundles in this column.  The scattering in the third column that is fully loaded with fuel is 1.6 $\sigma$ lower than expectation from a full cask.  While this level of disagreement has limited statistical significance, a slight deviation from the model of a full cask is expected, since some muon tracks which are projected here entered the cask and passed partially through the neighboring columns, which have less fuel content than the model of the full cask.

The fourth column, with a single missing bundle, displays an averaged scattering that is lower than the full cask model by 2.3$\sigma$, which indicates disagreement with expectations of the full cask at the 98$\%$ confidence level. Most of the relevant data for this column was taken in the position with both detectors centered on the cask over a period of 10 days.  If additional data were recorded for one month, the uncertainty would be decreased by a factor of two, providing much stronger constraints on the measured fuel content in this column.  In an actual verification scenario, this level of disagreement could be grounds for requesting additional measurements in this position, in order to draw stronger conclusions about the fuel content of this column.

\section{Conclusions}

We have shown that measurements of cosmic ray muon scattering can be used as a stand-alone method to independently determine if fuel assemblies are missing from a sealed dry storage cask.  Unlike more conventional radiographic probes, muons can penetrate the cask shielding and emerge with useful information on the cask contents.  Additionally, muons are an external probe that are not subject to backgrounds from other casks, and do not require any previous knowledge of the fuel history.  While statistical precision is limited by the natural flux of cosmic ray muons, count times on the order of weeks to several months can provide sufficient data to draw conclusions about cask content.  This potentially solves a long-standing problem in international nuclear safeguards, using charged particle tracking detectors and analysis techniques that are commonly found in high-energy particle physics.  

The 1.2 $\times$ 1.2 m$^2$ detectors used here are general purpose instruments that were not designed specifically for cask radiography, and cover a viewing area that is less than half the cask's  2.7 m diameter.  Due to the limited flux of cosmic ray muons and the multiple viewing positions required to survey the entire cask, $\sim$90 days of measurement time was required to record the data presented here.  Given that dry storage casks are designed to have a working lifetime of several decades, during which they typically sit undisturbed at storage facilities, several months of measurement time is not expected to pose a significant impediment to operations at commercial fuel storage sites.  This measurement time also satisfies the IAEA's timeliness goal for detecting the diversion of irradiated direct use material (such as plutonium in spent fuel) of 3 months \cite{IAEA_SG_glossary}.  However, an instrument which can cover more of the cask with a more efficient gamma rejection trigger can reduce count times by a factor of $\sim$4, and would allow multiple casks to be inspected at a site while still satisfying IAEA timeliness goals.  A dedicated instrument can also be further hardened against the environmental conditions (wind and precipitation) that are typically encountered at outdoor spent fuel storage installations.  Further work to understand the sensitivity of muon radiography to more complicated potential diversion scenarios (such as the removal of a portion of a single assembly, or replacement of a spent fuel assembly with a dummy assembly made of different material) is underway.

\section*{Acknowledgments}

We thank the personnel of the Idaho Nuclear Technology and Engineering Center at Idaho National Laboratory for their assistance.  This work is supported by the National Nuclear Security Administration's Office of Defense Nuclear Nonproliferation Research and Development.  This document is released under LA-UR-17-27060 and INL/JOU-17-43194.

\bibliography{casks}   

\end{document}